\documentclass[aps,prl,twocolumn,preprintnumbers,amsmath,amssymb,superscriptaddress]{revtex4-2}
\usepackage{graphicx}
\usepackage[caption=false]{subfig}
\usepackage{epsfig}
\usepackage{dcolumn}
\usepackage{bm}
\usepackage{color}
\usepackage{float}
\usepackage[colorlinks]{hyperref}
\usepackage{verbatim}
\usepackage{algorithm}
\usepackage{algorithmic}
\usepackage{tikz}
\usetikzlibrary{quantikz}

\hypersetup{citecolor = blue}

\def\be{\begin{equation}}       \def\ee{\end{equation}}
\def\bea{\begin{eqnarray}}      \def\eea{\end{eqnarray}}
\def\ba{\begin{array}}
\def\ea{\end{array}}
\def\bnum{\begin{enumerate} }
\def\enum{\end{enumerate}}

\def\=>{\Rightarrow}
\def\>{\rightarrow}

\def\eye2{Fathbb{I}}

\renewcommand{\c}[1]{{\cal #1}}

\newcommand{\no}{\nonumber}

\renewcommand{\>}{\rangle}

\renewcommand{\rm}[1]{\mathrm{#1}}

\usepackage{listings}
\usepackage{color}
\definecolor{lightgray}{gray}{1}

\begin{document}

\title{Probing many-body localization by excited-state VQE}

\author{Shuo Liu}
\affiliation{Institute for Advanced Study, Tsinghua University, Beijing 100084, China}
\affiliation{Tencent Quantum Laboratory, Tencent, Shenzhen, Guangdong 518057, China}
\author{Shi-Xin Zhang}
\email{shixinzhang@tencent.com}
\affiliation{Institute for Advanced Study, Tsinghua University, Beijing 100084, China}
\affiliation{Tencent Quantum Laboratory, Tencent, Shenzhen, Guangdong 518057, China}
\author{Chang-Yu Hsieh}
\email{kimhsieh@tencent.com}
\affiliation{Tencent Quantum Laboratory, Tencent, Shenzhen, Guangdong 518057, China}
\author{Shengyu Zhang}
\affiliation{Tencent Quantum Laboratory, Tencent, Shenzhen, Guangdong 518057, China}
\author{Hong Yao}
\email{yaohong@tsinghua.edu.cn}
\affiliation{Institute for Advanced Study, Tsinghua University, Beijing 100084, China}

\begin{abstract}
Non-equilibrium physics including many-body localization (MBL) has attracted increasing attentions, but theoretical approaches of reliably studying non-equilibrium properties remain quite limited. In this Letter, we propose a systematic approach to probe MBL phases 
via the excited-state variational quantum eigensolver (VQE) and demonstrate convincing results of MBL on a quantum hardware, which we believe paves a promising way for future simulations of non-equilibrium systems beyond the reach of classical computations in the noisy intermediate-scale quantum (NISQ) era. Moreover, the MBL probing protocol based on excited-state VQE is NISQ-friendly, as it can successfully differentiate the MBL phase from thermal phases with relatively-shallow quantum circuits, and it is also robust against the effect of quantum noises.
\end{abstract}

\date{\today}
\maketitle

{\bf Introduction:}
Many-body localization (MBL) is a novel dynamical phenomenon occurring in isolated many-body quantum systems. It has long been established that the quantum systems may enter MBL phases in the presence of sufficiently strong random disorder \cite{PhysRevLett.95.206603, basko_metal-insulator_2006, oganesyan_localization_2007, PhysRevB.75.155111, PhysRevB.77.064426, PhysRevB.81.134202, cuevas_level_2012, PhysRevLett.109.017202, PhysRevLett.110.067204, PhysRevLett.110.260601, PhysRevLett.111.127201, huse_phenomenology_2014} or quasi-periodic (QP) potential \cite{PhysRevB.87.134202, modak_many_2015, zhang_universal_2018, kohlert_observation_2019, zhang_strong_2019, ghosh_transport_2020} in one-dimensional (1D) systems. In MBL phases, the system fails to thermally equilibrate and exhibits 
exotic behaviors, such as ``area law'' entanglement for highly excited states \cite{PhysRevLett.110.067204, PhysRevLett.110.260601, PhysRevLett.111.127201}, logarithmic spread of entanglement \cite{PhysRevLett.109.017202}, and emergent local integrals of motion \cite{PhysRevLett.111.127201, huse_phenomenology_2014}. Such exotic non-equilibrium phases are qualitatively different from thermal phases that are often associated with the eigenstate thermalization hypothesis (ETH) \cite{PhysRevA.43.2046, PhysRevE.50.888, rigol_thermalization_2008, dalessio_quantum_2016, Deutsch_2018} and exhibit ``volume law'' entanglement in highly excited states \cite{Deutsch_2010, PhysRevX.8.021026}. However, various aspects of MBL remain elusive so far; for instance, whether MBL phases can survive in more than one dimensions is still under debate considering non-perturbative avalanche mechanism \cite{PhysRevB.95.155129, PhysRevLett.121.140601, PhysRevB.99.205149}.

To advance our understanding of non-equilibrium quantum phases, it is crucial to unambiguously detect and characterize possible MBL phases for various systems. Numerically, one can probe MBL by calculating the entanglement entropy or level statistics for eigenstates of a Hamiltonian via exact diagonalization, or directly simulating dynamical signatures such as charge imbalance or logarithmic entanglement spreading, following a quantum quench with time-evolution methods.
But, existing numerical approaches are often severely restricted due to the exponential scaling of Hilbert space as well as the excessively long evolution time required to simulate steady-state behaviors. Turning to experimental investigations, current hardware platforms still face many challenges, such as short coherence time and limited controllability for the Hamiltonian engineering. That whether MBL exists in more general systems, for example, higher dimensional systems \cite{choi_exploring_2016, bordia_coupling_2016, bordia_probing_2017, wahl_signatures_2019}, has not been rigorously established from current numerical simulations or analog-simulation based experiments. 

Quantum computer naturally helps as it can potentially simulate large quantum systems, beyond the capability of classical computers. A common approach is to directly simulate the time evolution of a quantum system using a quantum circuit and determine whether the system will thermalize after a long time by measuring dynamical observables such as charge imbalance. However, currently such idea is not NISQ-friendly since the required simulation time to confirm the existence of MBL can be much longer than the coherence time of the currently available quantum hardware. Namely, we are faced with the same challenge as other analog experiments without quantum error corrections. Instead, here we propose a more NISQ-friendly approach to detect MBL, relying less on quantum hardware resources and being more robust against noises in quantum hardware.

\begin{figure}[t]\centering
	\includegraphics[width=0.4\textwidth]{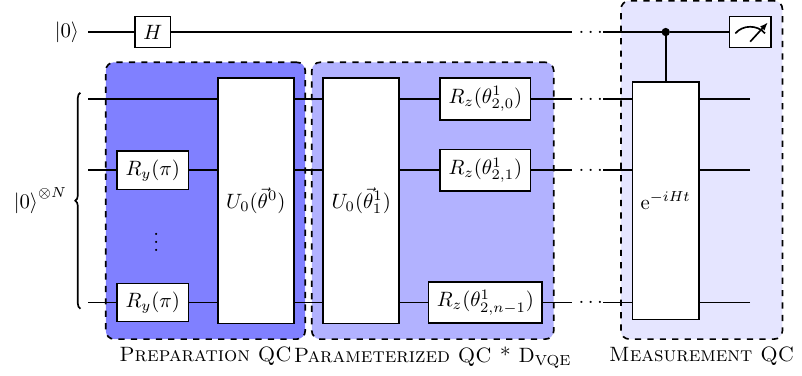}
	\caption{The circuit structure for the excited-state VQE and eigenstate witness measurement. In preparation circuit part, we first prepare an antiferromagnetic (AF) state $|0101...\rangle$ from the initial state $|0000...\rangle$ and then apply one layer of two-qubit entangling gates to generate the input state $|\psi_{0}\rangle$ for the PQC. The layer of two-qubit entangling gates is denoted as $U_{0}(\vec{\theta}^{0})$ and $U_{0}(\vec{\theta}^{0})$ = $ \prod_{j=0}^{N-1}\rm{exp}(\frac{i \pi \theta^{0}_{j} }{4}( \sigma^{x}_{j}\sigma^{x}_{j+1} + \sigma^{y}_{j}\sigma^{y}_{j+1}))$. We assume two-qubit entangling gates on different qubits sharing the same weight $\theta^{0}$ for simplicity.
	In the PQC part, each block of the ansatz is defined as $U(\vec{\theta}^{k})= (\prod_{j=0}^{N-1}\rm{Rz}(\theta^{k}_{2,j}))U_{0}(\vec{\theta}^{k}_{1})$, which respects $U(1)$ symmetry. And two-qubit entangling gates on different qubits have different weights. The depth of the PQC (number of blocks) $\rm{D}_{\rm{VQE}}$ can be adjusted.
}
	\label{fig:circuit}
\end{figure}

In this Letter, we propose a general MBL-probing protocol basing on variational quantum algorithms (VQAs). Recently, various quantum-classical hybrid variational algorithms, such as variational quantum eigensolver \cite{peruzzo_variational_2014} and quantum approximate optimization algorithm (QAOA) \cite{farhi_quantum_2014, McClean_2016,wang_quantum_2018, zhou_quantum_2020}, which are tailored for the NISQ hardware \cite{cerezo_variational_2020, bharti_noisy_2021, endo_hybrid_2021}, have been proposed. VQE is not only a representative VQA but also holds great potentials for near-term applications. For different practical problems encoded in VQE, we can define different objective functions (also known as cost functions). And the solutions are expected to give the minimum objective function. For ground state (lowest eigenstate) preparation for a Hamiltonian $H$, the objective function is defined as the expectation value of $H$. Firstly, we will generate a trial state $\vert \psi(\theta) \rangle = U(\theta) \vert \psi_{0} \rangle$ to approximate the solution, where $\vert \psi_{0} \rangle $ is a given initial state and $U(\theta)$ is an unitary matrix represented by the parameterized quantum circuit (PQC) as shown in Fig. \ref{fig:circuit}. When the PQC is deep enough, the ground state can be exactly written as $\vert \psi(\theta) \rangle$ for suitable parameters. Then, by updating the parameters in shallow PQC based on gradient descent, we will find a final converged state $\vert \psi(\theta) \rangle$ whose objective function can not decrease anymore and $\vert \psi(\theta) \rangle$ will be the approximate ground state. The VQE has been exploited in a variety of contexts from quantum chemistry \cite{omalley_scalable_2016, kandala_hardware-efficient_2017,  PhysRevX.8.011021, armaos_computational_2019, grimsley_adaptive_2019, tilly_computation_2020}, many-body physics \cite{PhysRevA.92.042303, dallaire-demers_low-depth_2018, 10.21468/SciPostPhys.6.3.029, McClean_2016, liu_variational_2019, cai_resource_2020, uvarov_variational_2020, tamiya_calculating_2021}, to lattice gauge theories \cite{PhysRevA.98.032331, kokail_self-verifying_2019}.

VQE can also be customized to search for excited states. There are several proposals utilizing VQE to discover low-lying excited states of quantum many-body systems, such as orthogonality constrained VQE \cite{PhysRevLett.87.167902, Cincio_2018, higgott_variational_2019, PhysRevA.99.062304}, which adds penalty projector terms to the Hamiltonian cost function that project out lower energy states, subspace expansion method \cite{PhysRevA.95.042308, nakanishi_subspace-search_2019}, which prepares variational states that span the low-energy manifold.
Furthermore, VQE can be adapted to search for highly excited states. The objective functions utilized in this case are either the energy variance $C(\theta) = \langle H^{2} \rangle - \langle H \rangle ^{2}$  \cite{bartlett_normal_1935, umrigar_optimized_1988, marotta_variational_2005, umrigar_energy_2005, khemani_obtaining_2016, pollmann_efficient_2016, vicentini_variational_2019, zhang_variational_2020, zhang_adaptive_2021}, which only vanishes for eigenstates of $H$ and is greater than zero for the superposition of different eigenstates of $H$ (see the Supplemental Material (SM) for details \footnote{See Supplemental Material at URL (to be added) for details, including the following: (1) a brief introduction to level spacing ratio for many-body localization transition, (2) the failure of energy variance to determine the VQE convergence, (3) EIPR and energy correlation between input states and output states of VQE, (4) scaling analysis of EIPR and eigenstate witness, (5) a brief introduction to eigenstate witness, (6) details of quantum noise and Trotter decomposition, (7) details for the real hardware experiments}), or $\langle (H-\lambda)^{2} \rangle$ \cite{macdonald_modified_1934, wang_solving_1994, peruzzo_variational_2014, mcclean_theory_2016, santagati_witnessing_2018}, which specifically targets an excited state closest to the energy $\lambda$. The many-body spectrum can be reconstructed by scanning through a range of $\lambda$ values within the energy width of $H$.

In this Letter, we propose to use the excited-state VQE with the energy-variance objective function to detect MBL phases. If the circuit ansatz is deep enough and with sufficient expressivity, then the converged state should be an eigenstate of the system. One can then measure the quantum state to infer purity or real-space inverse participation ratio (IPR) \cite{plerou_universal_1999, pradhan_correlations_2000} as the indicators of MBL phases. This option is straightforward, but it requires a (possibly exponentially) deep and coherent quantum circuit beyond the NISQ regime. We propose an alternative approach, where only the excited-state VQE with a NISQ compatible ansatz is required. In this case, the converged state is not necessarily an eigenstate, especially on the thermal side, where the eigenstate manifests ``volume law'' entanglement and may not be fully represented with a shallow circuit. Therefore, the indicator to differentiate between MBL and ergodic phases is replaced with the converged performance of the excited-state VQE. On the MBL side, the final state is more quickly converged in terms of the eigenspace distribution. Accordingly, we propose an experimentally measurable quantity that could help assess a state's convergence in the eigenspace. Since our proposed method does not rely on any properties of a particular model, it can be readily applied to a broad range of systems.

{\bf Model:}
We illustrate the proposed method by investigating the interacting Aubry-Andr\'e (AA) model \cite{aubry1980analyticity, lahini_direct_2009, biddle_localization_2009, modak_many_2015, dutta_many-body_2018, kohlert_observation_2019}, a well-studied system hosting the many-body localization transition. The Hamiltonian of the interacting AA model reads:
\begin{eqnarray}
&&H=\sum_{i}(\sigma^{x}_{i}\sigma^{x}_{i+1}+\sigma^{y}_{i}\sigma^{y}_{i+1} +V_{0}\sigma_{i}^{z} \sigma_{i+1}^{z})\no\\
&&~~~~~~+W \sum_{i=1}^N \cos(2 \pi \eta i+\phi) \sigma_{i}^{z},
\end{eqnarray}
where $W$ is the strength of quasi-periodic potential, $\sigma^\alpha$ are Pauli matrices, $N$ is the size of the system, and $\phi$ is the phase of the cosine potential. We set $V_{0}=0.5$ and $\eta =(\sqrt{5}-1)/2$ throughout the work. We can numerically determine the MBL transition point via the level spacing ratios for this model \cite{oganesyan_localization_2007, throckmorton_studying_2021} (see the SM for details \cite{Note1}).

{\bf Circuit ansatz:}
Our proposed circuit ansatz for the excited-state VQE consists of two parts: the input-state preparation circuit and the parameterized quantum circuit, acting as the variational ansatz to optimize the cost function, as shown in Fig. \ref{fig:circuit}. Since the AA model conserves the total spin polarization along the $z$ direction, we focus on the total spin $M_{z}=\sum_i \sigma_i^z=0$ sector. Namely, the quantum gates employed in the circuit ansatz should respect this $U(1)$ symmetry. The cost function for the excited-state VQE is the energy variance $\langle H^{2} \rangle-  \langle H \rangle ^{2}$, as mentioned before.

\begin{figure}[t]\centering
	\includegraphics[width=0.42\textwidth]{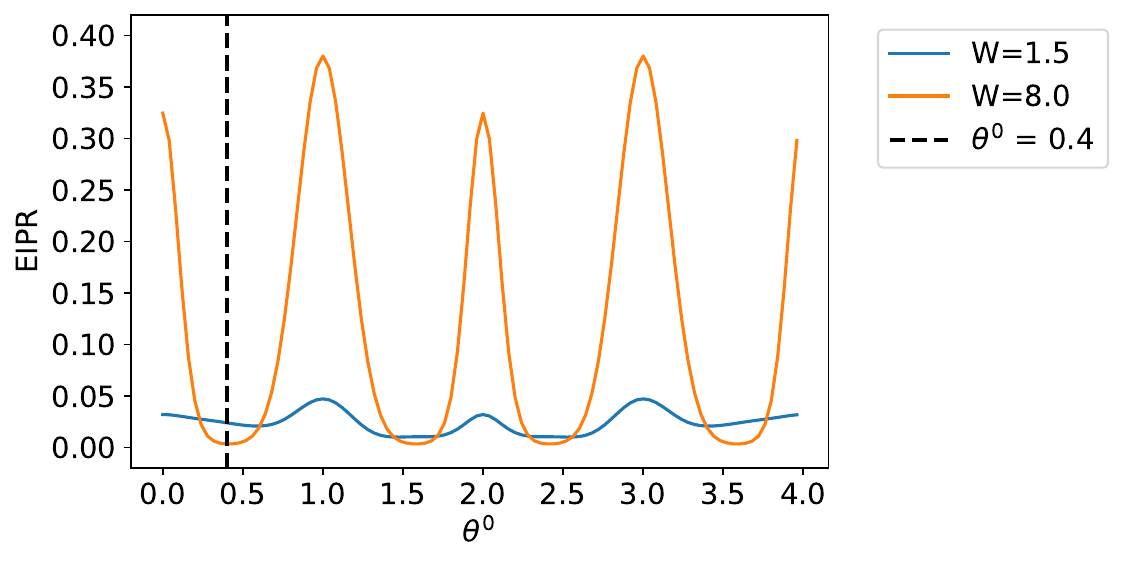}
	\caption{EIPR of the input state $\vert \psi_0\rangle$ obtained from the preparation circuit for $W=1.5$ and $8.0$, respectively. Here $N=12$.}
	\label{fig:eipr}
\end{figure}

{\bf Eigenspace inverse participation ratio:}
The final converged state $\vert\psi \rangle$ obtained from the excited-state VQE is a superposition of eigenstates, localized within an energy window, which may not be easily resolved by the energy variance indicator because of the extremely small energy separation between adjacent energy levels (see the SM for details \cite{Note1}). Theoretically, we may use the eigenspace inverse participation ratio (EIPR) \cite{santagati_witnessing_2018} to gauge the extent of convergence onto an eigenstate under the excited-state VQE minimization:
\begin{eqnarray}
\rm{EIPR}(\vert\psi\rangle) = \sum_{n} \vert\psi_{n}\vert^{4},
\end{eqnarray}
where $\vert\psi \rangle = \sum_{n} \psi_{n} \vert n\rangle$ and $H\vert n\rangle = \lambda_{n} \vert n\rangle$ with $\vert n\rangle$ is the $n$-{th} eigenstate of the Hamiltonian $H$. For each exact eigenstate of $H$, its EIPR is one. For a state which is a linear superposition of different eigenstates, its EIPR is less than one. 
For the maximally randomized state in eigenspace $\vert\psi\rangle=\sum_{n}\frac{1}{\sqrt{2^N}}\vert n\rangle$, where $2^N$ is the dimension of the Hilbert space of the system; $\rm{EIPR}=1/2^N\rightarrow 0$ as $N\rightarrow \infty$ in the thermodynamic limit. Consequently, EIPR is capable in efficiently determining whether the final converged state is exactly an eigenstate of the Hamiltonian $H$ or a superposition of multiple eigenstates.

We firstly investigate the EIPR for the input state $\vert \psi_0\rangle$, created with the preparation circuit of Fig. \ref{fig:circuit}. The results are shown in Fig. \ref{fig:eipr}. When the system is in the MBL phase ($W=8.0$), some of the shallow-circuit input states (adjusted by the parameter $\theta^0$) may achieve high EIPR, implying the ``area law'' entanglement \cite{PhysRevLett.109.017202, PhysRevLett.110.067204, PhysRevLett.110.260601, PhysRevLett.111.127201} for highly-excited eigenstates. On the other hand, when the system is in the thermal phase ($W=1.5$), all input states consistently score low EIPR, indicating that a shallow circuit has difficulty to approximate the system's eigenstates. Overall, the results shown in Fig. \ref{fig:eipr} are consistent with the understanding of the AA model. Furthermore, we find a positive correlation between EIPR of input states and EIPR of output states, as well as a positive correlation between energy of input states and energy of output states (see the SM for details \cite{Note1}). For the following investigations, we then fix $\theta^{0}=0.4$ for input-state preparation. As shown in Fig. \ref{fig:eipr}, the MBL phase has a smaller initial EIPR and the input state's energy is in the middle of energy spectrum in both phases. By this choice, under the excited-state VQE, the output states will likely to converge to some highly excited states, and a high EIPR for output state in MBL phase can be reliably attributed to the nature of the phase instead of a purely better optimization start point.

\begin{figure}[t]\centering
	\includegraphics[width=0.4\textwidth]{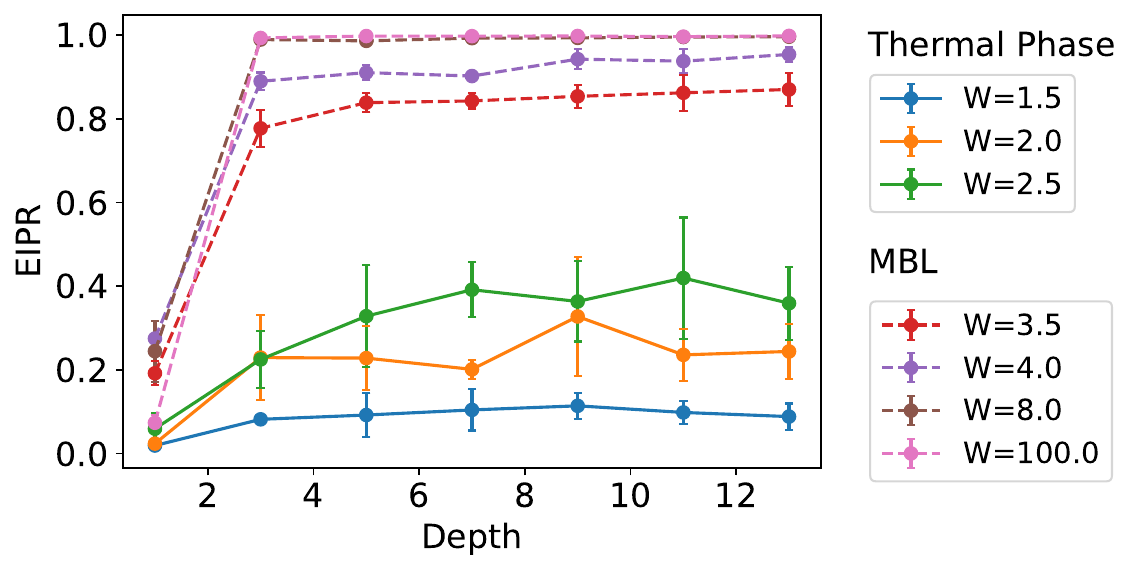}
	\caption{EIPR of final converged states from excited-state VQE on 12-qubit system with different potential strength $W$.}
	\label{fig:eipr_final}
\end{figure}

With the noiseless quantum circuit simulation \cite{zhang_tensorcircuit_2022}, we calculate EIPR for converged states with varying PQC depth. Theoretically, the expressivity of the variational ansatz increases with the circuit depth, and the achieved EIPR should also improve correspondingly. Indeed, the results for converged states' EIPR, as shown in Fig. \ref{fig:eipr_final}, support this intuition. Although the initial EIPR of the MBL phase is smaller than that of the thermal phase and the optimization starts from the same input state, the final EIPR of the MBL phase is much larger and extremely close to $1$ even with a relatively shallow quantum circuit. There is a clear gap of EIPR between thermal and MBL phases, and the gap grows with system size (see the SM for details \cite{Note1}). These results confirm that excited-state VQE with shallow circuits indeed performs qualitatively better for MBL systems.

\begin{figure}[t]\centering
	\includegraphics[width=0.50\textwidth]{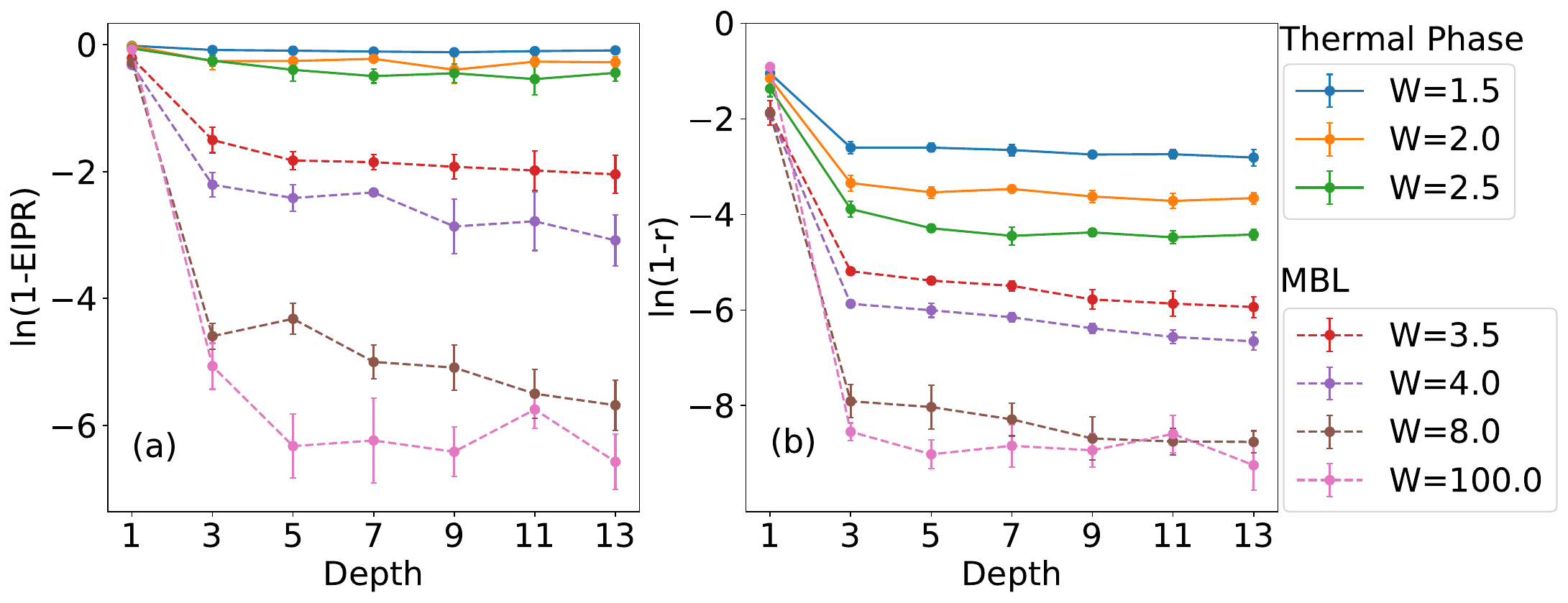}
	\caption{(a), $\ln(1-\rm{EIPR})$ of final converged states. (b), $\ln(1-r)$ of final converged states, where $r$ is the eigenstate witness and experimentally accessible. Here $N=12$.}
	\label{fig:lnr_eipr}
\end{figure}

{\bf Experimental relevance:}
Though EIPR of the output states significantly differs between MBL and thermal phases in numerical simulations, it can not be directly observed in experiments. To determine whether a state is sufficiently enough to an eigenstate, an experimentally accessible proxy is desired. To this end,
the eigenstate witness \cite{santagati_witnessing_2018} is such a quantity which can be experimentally determined via purity measurements. The circuit for eigenstate witness measurement is shown in Fig. \ref{fig:circuit}.

We illustrate the notation of the eigenstate witness 
as follows. The output state obtained from the excited-state VQE can be written as $\vert\psi\rangle =\sum_{n}\alpha_{n} \vert n\rangle$, where $\vert n\rangle$ represents eigenstate of $H$ with eigenenergy $\lambda_n$. With an extra ancilla qubit initialized in a superposition state $\vert + \rangle=(\vert 0\rangle+\vert 1\rangle)/\sqrt{2}$, after a controlled time evolution under $H$, the reduced density matrix for the ancilla qubit reads:
\begin{eqnarray}
\rho_{\rm{reduced}}=\left(	
\begin{matrix}
\frac{1}{2} & \frac{1}{2}\sum_{n}\vert\alpha_{n}\vert^{2}\rm{e}^{i\lambda_{n}t}\\
\frac{1}{2}\sum_{n}\vert\alpha_{n}\vert^{2}\rm{e}^{-i\lambda_{n}t} & \frac{1}{2}
\end{matrix}
\right),
\end{eqnarray}
and the eigenstate witness is defined as the purity of the ancilla qubit:
$
r = \rm{Tr}(\rho_{\rm{reduced}}^{2}),
$
which can be estimated using randomized measurements in experiments \cite{PhysRevA.99.052323, PhysRevLett.124.010504, haug_large-scale_2021} or a simple state tomography. Such witness is lower bounded by the corresponding EIPR of the output states and can reflect the convergence performance of the excited-state VQE (see the SM for details \cite{Note1}).

We choose the evolution time $t=1.0/W$, as the bandwidth is roughly proportional to $W$. The results of $\rm{ln}(1-r)$ are shown in Fig. \ref{fig:lnr_eipr}. For each data point in Fig. \ref{fig:lnr_eipr}, the number is averaged over 10 best out of 100 independent optimization results for the excited-state VQE. Both EIPR and $r$ of the MBL phase are larger than those of the thermal phase. These differences will be more prominent as the system size increases.

Theoretically, $\ln(1-r)$ should approach the minus infinity if the output state is exactly an eigenstate of the system. However, due to the effectiveness of an optimization routine and numerical accuracy of the simulation, the $(1-r)$ value may only reach an ``effective zero'' from above. In Fig. \ref{fig:lnr_eipr}, $(1-\rm{EIPR})$ and $(1-r)$ have this ``effective zero'' which is roughly determined by the data from $W=100.0$, supposedly in a deep region of the MBL phase. The effective zero of $(1-r)$ is around $-8.9$ in the log scale. And the circuit depth required to reach this effective zero depends on Hamiltonian parameter $W$. We estimate the required depth in Table. \ref{tab:depth} (the required depth is fitted by assuming linear relation between $\ln(1-r)$ and the circuit depth). On the thermal side, the depth required for perfect excited-state convergence is close to $2^{N}$; while on the MBL side, the depth required is in the order of $O(N)$.
The depths needed to reach effective zero can serve as an another indicator differentiating between the MBL and thermal phases in systems whose phases are unknown. In particular, it is easier to distinguish the two phases when the system size increases.

\begin{table}[b]
		\centering
		\begin{tabular}{c|c|c|c|c|c|c}\hline
			~~~~~W~~~~~ & ~~~1.5~~~ & ~~~2.0~~~ & ~~~2.5~~~ & ~~~3.5~~~ & ~~~4.0~~~ & ~~~8.0~~~ \\\hline
			Depths & 245 & 197 & 256 & 52 & 38 & 14 \\\hline
		\end{tabular}
		\caption{Fitted circuit depths required to perfectly converge to the excited state for different $W$.}
		\label{tab:depth}
\end{table}

\begin{figure}[t]\centering
	\includegraphics[width=0.50\textwidth]{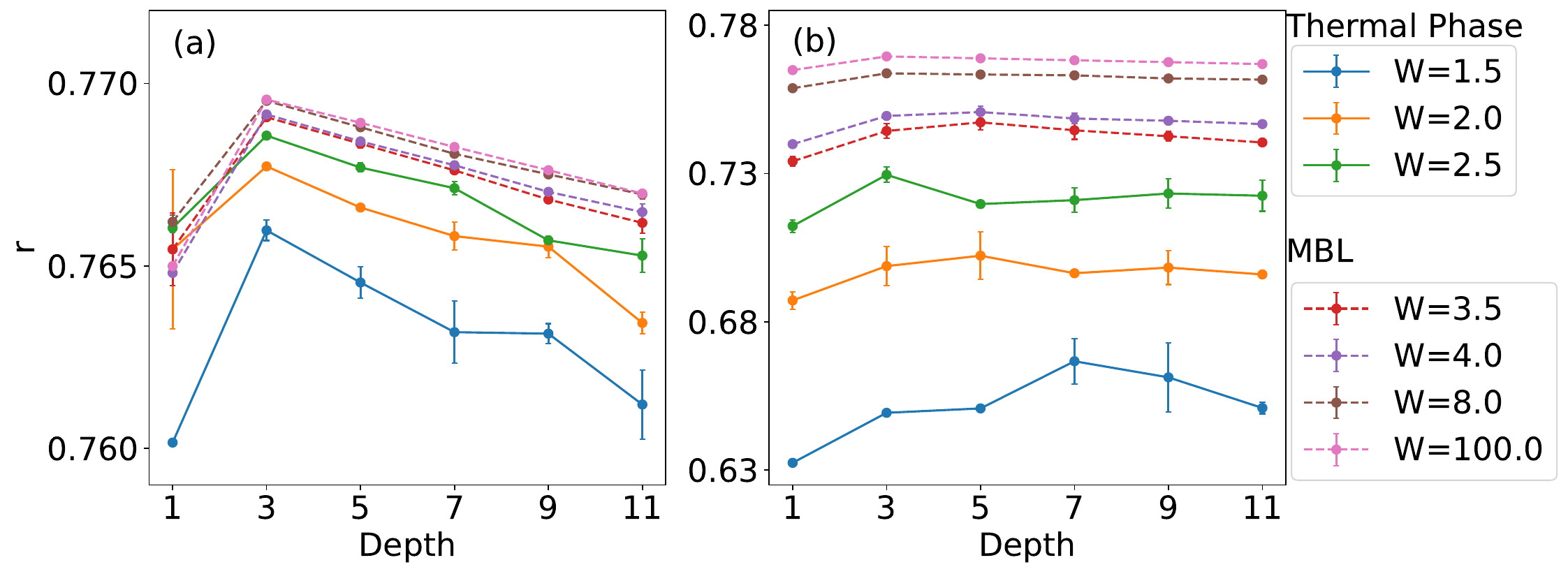}
	\caption{(a), $r$ of final converged states from a noisy simulation without Trotter decomposition. (b), $r$ of final converged states from a noisy simulation with one time slice Trotter decomposition. Here $N=8$.}
	\label{fig:r_noise_trotter}
\end{figure}

{\bf Effect of quantum noise:}
Our approach of probing MBL phases has potential quantum advantage as it can be extended to larger systems and higher dimensions in principle. However, for current quantum computing hardware, we must consider the effects of noise which may compromise the performance of excited-state VQE. One important question is whether we can apply our method to a real device with noises. To answer this question, we carry out similar computation in the presence of a quantum depolarizing channel with noise strength $p=10^{-3}$ after each two-qubit gate. The results are averaged over $5$ best of $20$ independent trials on an 8-qubit system (the results for larger size systems with the presence of quantum noise obtained by the Monte Carlo trajectory method can be found in the Supplemental Material).
As indicated in Fig. \ref{fig:r_noise_trotter}(a), $r$ does not keep increasing with PQC depths any more; instead, it could decrease due to the accumulated noisy effects for deeper PQC. Nonetheless, the eigenstate witness $r$ behaves sufficiently distinct in the two phases. The magnitude of $r$ from the MBL side is still significantly larger than that from the thermal side. Besides, $r$ shows better noise resilience when the system enters the MBL side.  In the MBL phase, local perturbations spread only logarithmically in time \cite{RevModPhys.91.021001}, as opposed to the algebraic spreading in thermalizing dynamics. Therefore, $r$ of the thermal phase gets more severely affected due to the faster spreading of quantum noise.

When implementing the controlled time evolution in a real device, one also needs to consider the error brought by the Trotter decomposition. Here we design a type of decomposition that reduces the number of two-qubit gates, and we propagate the system with only one time slice to further reduce the total circuit depth for the digital quantum simulation (see the SM for details \cite{Note1}). The eigenstate witness $r$, after incorporating the effects of both noise and Trotter decomposition, is shown in Fig. \ref{fig:r_noise_trotter}(b), which still behaves distinctly in the two phases. The numerical results confirm that our proposal is NISQ-friendly and ready to be validated in a quantum device.

\begin{figure}[t]\centering
	\includegraphics[width=0.48\textwidth]{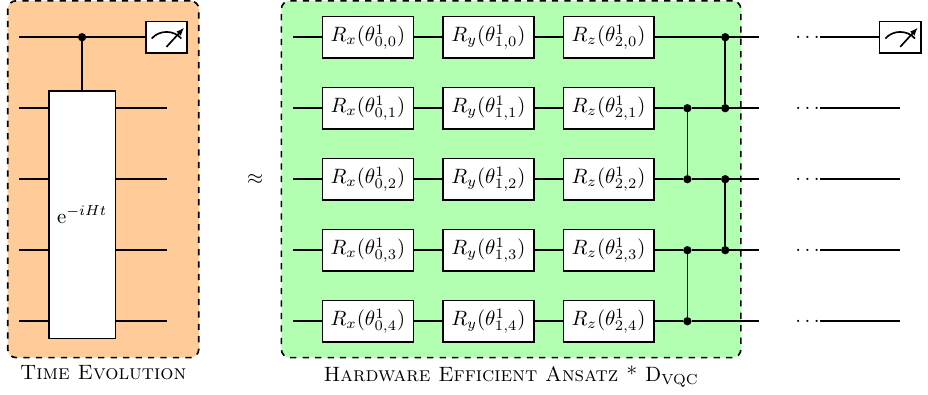}
	\caption{VQC using hardware-efficient ansatz to approximate controlled time evolution. The two-qubit entangling gate is $\rm{CZ}$ gate.}
	\label{fig:vqc}
\end{figure}

{\bf Real Hardware Experiments:}
We apply our method to a four-site model on the available IBM open access quantum hardware. 
Since the controlled time evolution module is still expensive to implement on NISQ devices \cite{PhysRevResearch.1.013006, PhysRevA.104.042418}, we further reduce the quantum resources by utilizing a variational quantum circuit (see Fig. \ref{fig:vqc}) to approximate the controlled time evolution module. The VQC is optimized via the normal VQE classically with following cost function:
\begin{eqnarray}
C(\theta) = -\frac{ \rm{Tr}(U(\theta)V^{\dagger})}{2^{N}},
\end{eqnarray}
where $U(\theta)$ is the unitary ansatz of VQC and $N$ is the number of qubits, and
\begin{eqnarray}
V =\left(	
\begin{matrix}
I     & 0\\
0     & \rm{e}^{iHt}
\end{matrix}
\right),
\end{eqnarray}
with the ancilla qubit initialized to $\vert + \rangle=(\vert 0\rangle+\vert 1\rangle)/\sqrt{2}$ state.
Because $U(\theta)V^{\dagger}$ is also a unitary matrix of dimension $2^{N}$, the cost function has a minimum of $-1$ when $U(\theta)=V$. We can use $U(\theta)$ as the ansatz to approximate the controlled time evolution in the real hardware. Such a variational circuit is constructed via the hardware-efficient ansatz and thus has less number of two-qubit gates in total while maintaining a high fidelity against the exact controlled time evolution (see the Supplemental Material for more detail). For large system sizes, the loss function defined above is hard to simulate. However, the unitary $U$ to approximate the controlled time evolution module $V$ can also be obtained based on quantum-assisted quantum compiling (QAQC) algorithm with an alternative loss function using the Local Hilbert-Schmit Test (LHST) \cite{Khatri2019quantumassisted}, which has better scalability and less barren plateau effect. Therefore, we can carry out the optimization on real quantum hardware when the system size is too large to simulate in silico, i.e. we trade off the depth of the time evolution circuit with a large number of shallower variational circuit execution on real devices to make our scheme more suitable on NISQ devices.

\begin{figure}[t]\centering
	\includegraphics[width=0.4\textwidth]{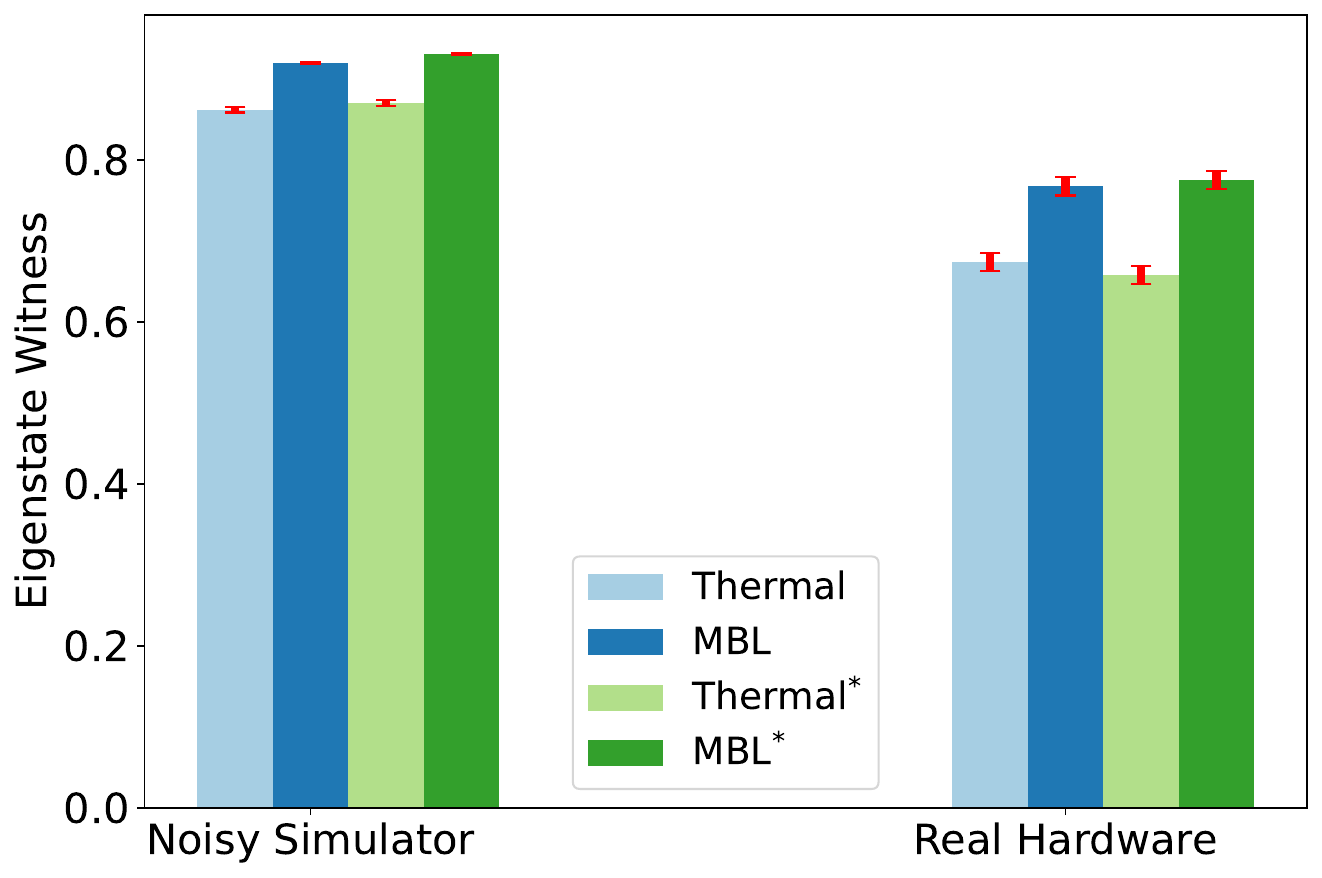}
	\caption{Eigenstate witness measured on the noisy simulator and real hardware (IBM\_Santiago) (in the inset $^{*}$ means the results are processed with readout error mitigation). The estimation error bars due to finite shots of measurement are also included.}
	\label{fig:resultibm}
\end{figure}

We demonstrate our method in a noisy hardware simulator based on IBMQ hardware device (specifically we use the IBM\_Santiago instance). We set $W=8.0$ for the MBL phase and $W=1.5$ for the thermal phase, and we choose the evolved time $t=0.15/W$. The parameters of the excited-state VQE and VQC are both determined by the best of $20$ independent VQE trials. In Fig. \ref{fig:resultibm}, the results of the noisy simulator are the average of $100*8192$ independent measurement shots and the results of real hardware are the average of $50*8192$ independent measurement shots.
A qualitative difference is clearly seen which is sufficient to distinguish the two phases. In fact, for this case, the difference obtained in quantum hardware experiments is even more evident than that obtained from the noisy simulation. 

{\bf Discussions and concluding remarks:}
We have demonstrated that the excited-state VQE with EIPR constitutes a reliable method to probe MBL phase. Since this proposed method requires only a shallow circuit, it can be 
executed on NISQ hardware when we substitute the EIPR with the eigenstate witness protocol. It is worth noting that though we can use a variational quantum circuit to approximate the controlled time evolution module in principle, this approximation is still challenging on real NISQ device due to the variational optimization methodology and the depth required for the variational circuit replacement to achieve satisfying fidelity. Still, compared to the conventional approaches that characterize the MBL phase with late-time dynamics, the evolution time required in our method is much shorter and is more compatible with the current-generation quantum computers.

There are many promising directions for further studies. In the present work we have considered only one circuit ansatz. Other different PQC ansatz for excited-state VQE can also be systematically examined for potential benefits. The circuit ansatz with better expressiveness may characterize the highly excited states better, and the optimal ansatz can be automatically designed via techniques from quantum architecture search \cite{zhang_differentiable_2020, zhang_neural_2021}. We can also utilize post-processing enhancements \cite{zhang_variational_2021} which hopefully improve the performance of excited-state VQE. As our method is independent of the microscopic details of the model considered in this work, it can be straightforwardly extended to investigate other models which might host many-body localization transitions. Moreover, our method can hopefully investigate whether MBL phases can survive in higher dimensions in the future, once the quantum hardware can offer longer coherence time.

~\newline
\textbf{Acknowledgements:} We thank Z.-Q. Wan for helpful discussions. This work is supported in part by the Beijing Natural Science Foundation under Grant No. Z180010
(H.Y.), the NSFC under Grant No. 11825404 (S.-X.Z., S.L., and H.Y.), the CAS Strategic Priority Research Program under Grant No. XDB28000000 (H.Y.), and Beijing Municipal Science and Technology Commission under Grant No.Z181100004218001 (H.Y.).

\bibliographystyle{apsreve}
\bibliography{ref}

\clearpage

\begin{widetext}
	\section*{Supplemental Materials}
	\renewcommand{\theequation}{S\arabic{equation}}
	\setcounter{equation}{0}
	\renewcommand{\thefigure}{S\arabic{figure}}
	\setcounter{figure}{0}

	\subsection{A. Level spacing ratio for many-body localization transition }
	The level spacing ratio is defined as:
	\begin{eqnarray}
	r_{\phi}=\frac{\rm{min}\{\Delta^{n}_{\phi},\Delta^{n+1}_{\phi}\}}{\rm{max}\{\Delta^{n}_{\phi},\Delta^{n+1}_{\phi}\}},
	\end{eqnarray}
	where $\Delta^{n}_{\phi} = E^{n}_{\phi}-E^{n+1}_{\phi}$ is the gap between eigenenergy levels $n$ and $n+1$, and $\phi$ labels a given quasi-periodic potential, $[r^{(n)}_{\phi}]$ is the level average of this ratio. In the delocalized phase (thermal phase), the level spacings follow a Gaussian orthogonal ensemble (GOE) distribution and the level-averaged level spacing ratio is $[r^{(n)}_{\phi}] \approx 0.53$ \cite{PhysRevLett.110.084101}. In contrast, in the localized phase, the level spacing follows a Poisson distribution, which gives $[r^{(n)}_{\phi}] = 2 \ln2 - 1 \approx 0.39$ \cite{PhysRevLett.110.084101}. The results of level spacing ratios for the interacting AA model are given in Fig. \ref{fig:r}.

	\begin{figure}[H]\centering
		\includegraphics[width=0.40\textwidth]{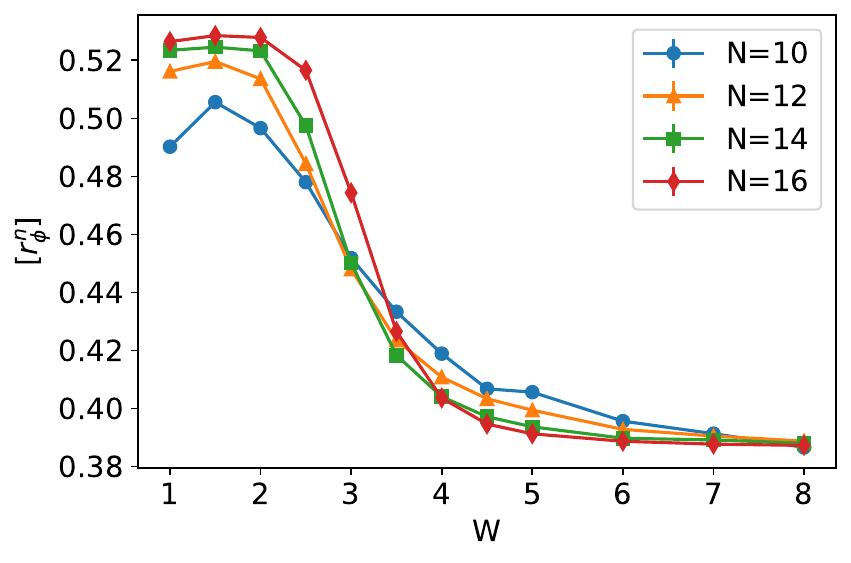}
		\caption{Exact diagonalization calculation: level spacing ratio with varying quasi-periodic potential strength $W$. Each point is an average of 1000 different random phases $\phi$.}
		\label{fig:r}
	\end{figure}	

	\subsection{B. The failure of energy variance to determine the VQE convergence}
	Consider a state $\vert \psi \rangle$:
	\begin{eqnarray}
	\vert \psi \rangle = \sum_{n} \psi_{n} \vert n \rangle,
	\end{eqnarray}
	where $\vert n \rangle $ is the $n$-th eigenstate of the Hamiltonian $H$ and $H \vert n \rangle = \lambda_{n} \vert n \rangle$. The energy variance is
	\begin{eqnarray}
	\langle \psi \vert H^2 \vert \psi \rangle - \langle \psi \vert H \vert \psi \rangle^2 &=& \sum_{m, n} \psi^{*}_{m} \psi_{n}\langle m \vert H^2 \vert n \rangle - (\sum_{m,n} \psi^{*}_{m} \psi_{n}\langle m \vert H \vert n \rangle)^2 \\ \nonumber
	&=& \sum_{n} \vert \psi_{n} \vert^2 \lambda_{n}^2 - (\sum_{n}  \vert \psi_{n} \vert^2 \lambda_{n})^2 \\ \nonumber
	&=& (\sum_{m} \vert \psi_{m} \vert^2) (\sum_{n} \vert \psi_{n} \vert^2 \lambda_{n}^2) - (\sum_{n}  \vert \psi_{n} \vert^2 \lambda_{n})^2 \\ \nonumber
    &\ge& (\sum_{n}  \vert \psi_{n} \vert^2 \lambda_{n})^2 - (\sum_{n}  \vert \psi_{n} \vert^2 \lambda_{n})^2 \\ \nonumber
    &=& 0,
	\end{eqnarray}
	where we have utilized the Cauchy–Schwarz inequality. When $\vert \psi \rangle$ is an eigenstate of $H$, i.e., $\vert \psi \rangle = \vert n \rangle$, the energy variance is zero. Otherwise, the energy variance is positive. Therefore, the energy variance can be used as the cost function for the excited-state VQE. 

	The final converged state $\vert \psi \rangle$ obtained from the excited-state VQE is a superposition of eigenstates, mainly localized within an energy window,
    which may not be easily resolved by the energy variance indicator because of the extremely small energy separation between adjacent energy levels. Assume the final converged state is simply a superposition of two eigenstates: $\vert\psi\rangle=a\vert E \rangle+b\vert E+\Delta\rangle$, where $a^{2}+b^{2}=1$. $\vert E \rangle$ and $\vert E+\Delta\rangle$ are eigenstates with energy $E$ and $E+\Delta$, respectively, and $\Delta$ is the energy gap which is extremely close to zero. Then
	\begin{eqnarray}
	\langle H \rangle=\langle\psi\vert H \vert\psi\rangle&=&a^{2}E+b^{2}(E+\Delta)\\ \no
    &=&E+(1-a^{2})\Delta,
	\end{eqnarray}
	and
	\begin{eqnarray}
	\langle H^{2}\rangle=\langle\psi\vert H^{2}\vert\psi \rangle&=&a^{2}E^{2}+b^{2}(E+\Delta)^{2}\\ \no
	&=&E^{2}+(1-a^{2})(2E\Delta+\Delta^{2}).
	\end{eqnarray}
	The energy variance is:
	\begin{eqnarray}
	\langle H^{2}\rangle - \langle H\rangle^{2} =(a^{2}-a^{4})\Delta^{2}.
	\end{eqnarray}
	It will approach zero even if the final state is a superposition of different eigenstates, as long as $\Delta$ is close to zero. In other words, the energy variance is no longer a good indicator to verify whether the excited-state VQE converged to one eigenstate in the limit of small $\Delta$.

	\subsection{C. The EIPR and energy correlation between input states and output states of VQE}
	The initial input state $\vert\psi_{0}\rangle$ is a superposition of eigenstates. If the initial EIPR is large, which means one eigenstate has already been dominant since the beginning, then the final converged state for a VQE calculation shall flow to that eigenstate. As shown in Fig. \ref{fig:eipr_SM}, EIPR is maximum when $\theta^{0}=1.0$ and $W=8.0$. We show the spectral decomposition of this initial input state in the eigenspace in Fig. \ref{fig:overlap_25_9}. The EIPR of the final output state is almost $1$ and successfully converges to that dominant eigenstate as shown in Fig. \ref{fig:overlap_25_9}. Even if we randomize the initial parameters of PQC in a small range, then all final converged states will still flow to that eigenstate as shown in Fig. \ref{fig:final_25}. When the initial EIPR is small, the input state is more uniformly distributed in the eigenspace as shown in Fig. \ref{fig:overlap_10_9}. In such cases, the performance of the excited-state VQE deteriorates, the final converged state is stuck as a superposition of different eigenstates. Furthermore, the dominant eigenstate of the output state will change in different independent trials as shown in Fig. \ref{fig:final_10}.

	\begin{figure}[t]\centering
	\includegraphics[width=0.48\textwidth]{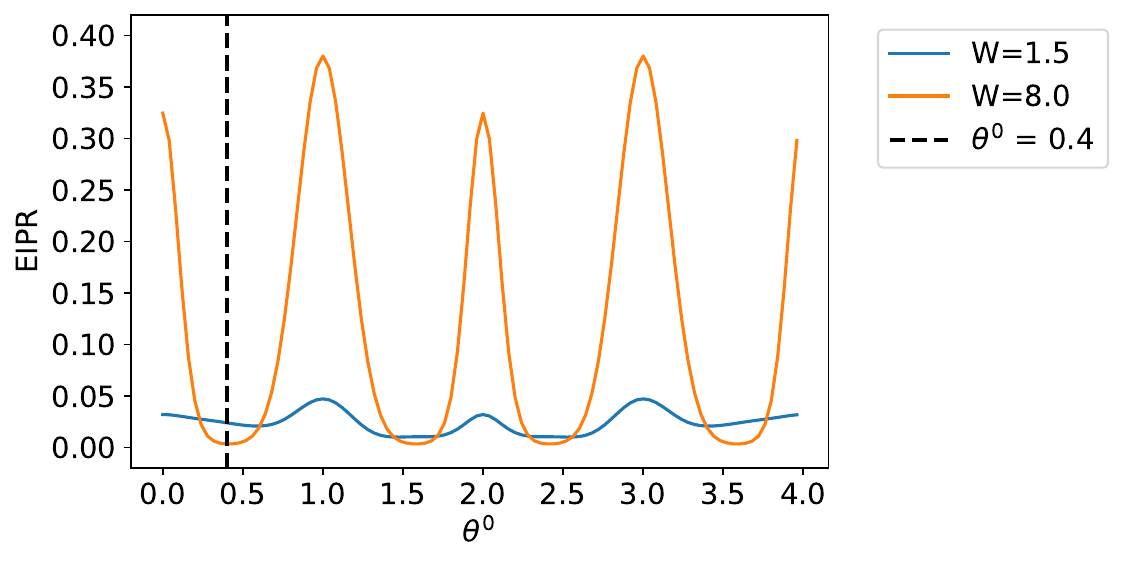}
	\caption{EIPR of input state $\vert \psi_0\rangle$ determined by the preparation circuit with $W=1.5,8.0$; $N=12$.}
	\label{fig:eipr_SM}
	\end{figure}

    \begin{figure}[t]\centering
		\includegraphics[width=0.9\textwidth]{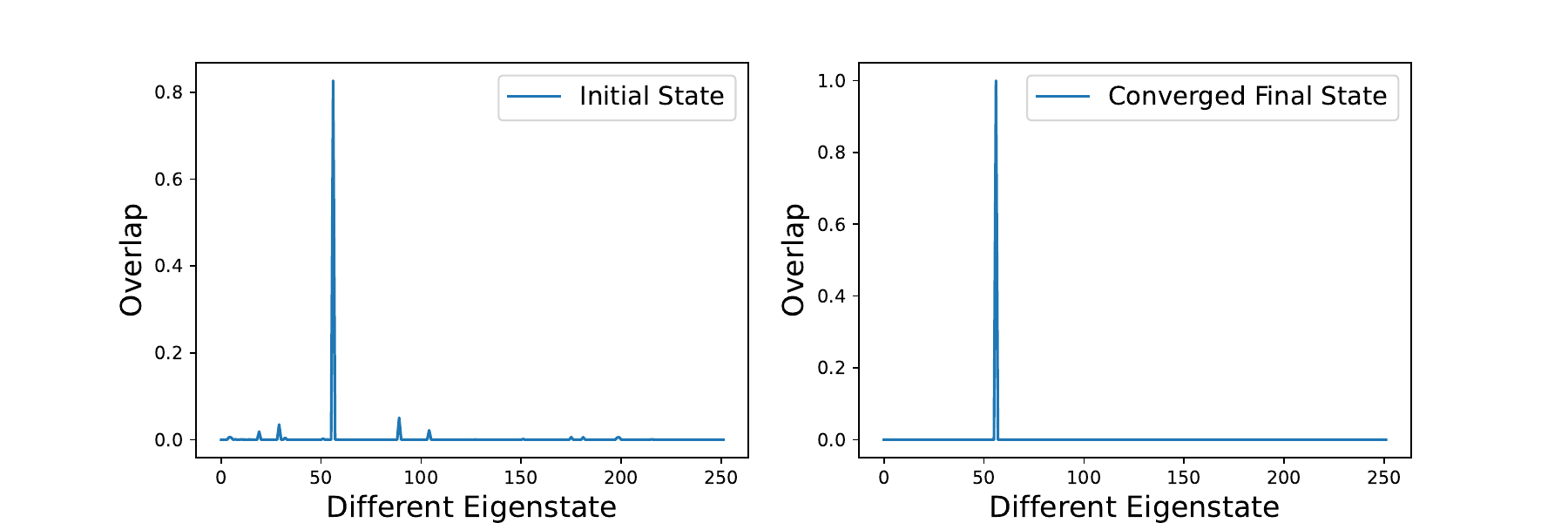}
		\caption{Overlap with different eigenstates: input state vs. final converged state: $W=8.0$; $N=10$; depth=3.}
		\label{fig:overlap_25_9}
	\end{figure}

	\begin{figure}[t]\centering
		\includegraphics[width=0.40\textwidth]{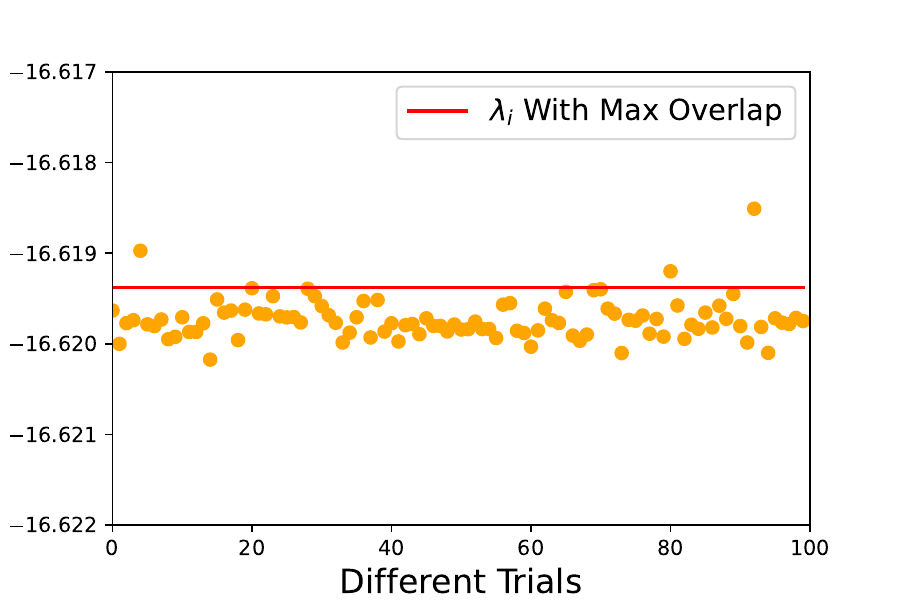}
		\caption{Final converged energy for $W=8.0$ and input state of good $\theta^{0}$; $N=10$; depth=3.}
		\label{fig:final_25}
	\end{figure}

	\begin{figure}[t]\centering
		\includegraphics[width=0.9\textwidth]{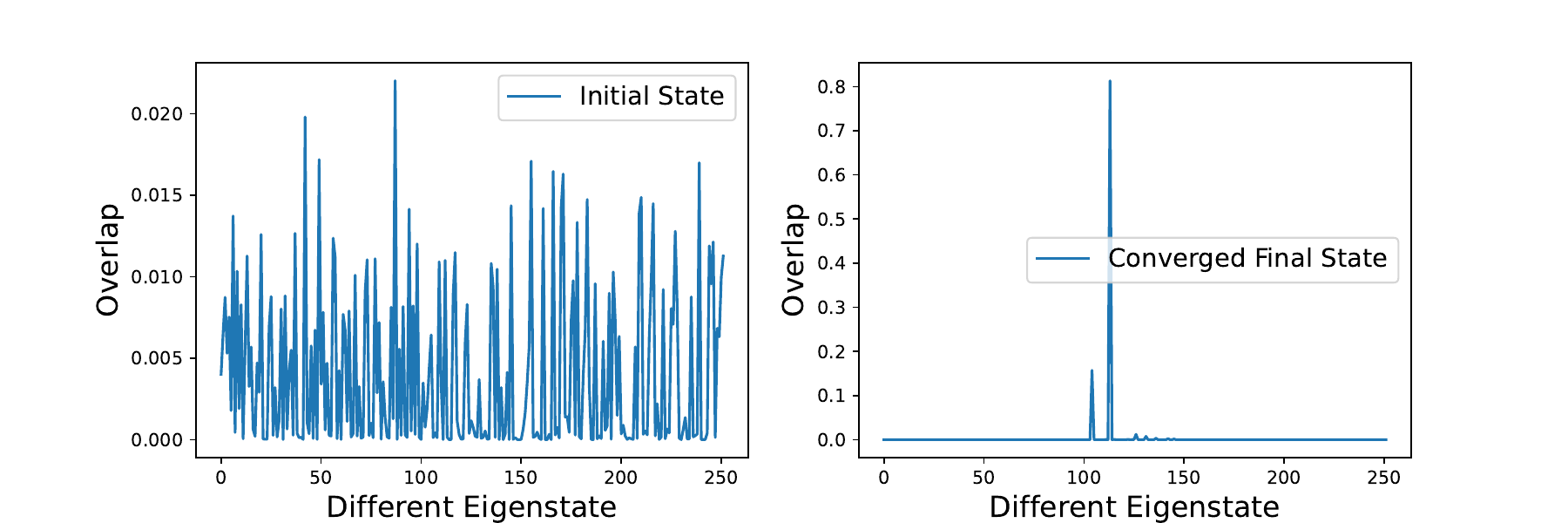}
		\caption{Overlap with different eigenstates: input state vs. final converged state: $W=8.0$; $N=10$; depth=3.}
		\label{fig:overlap_10_9}
	\end{figure}

	\begin{figure}[t]\centering
		\includegraphics[width=0.40\textwidth]{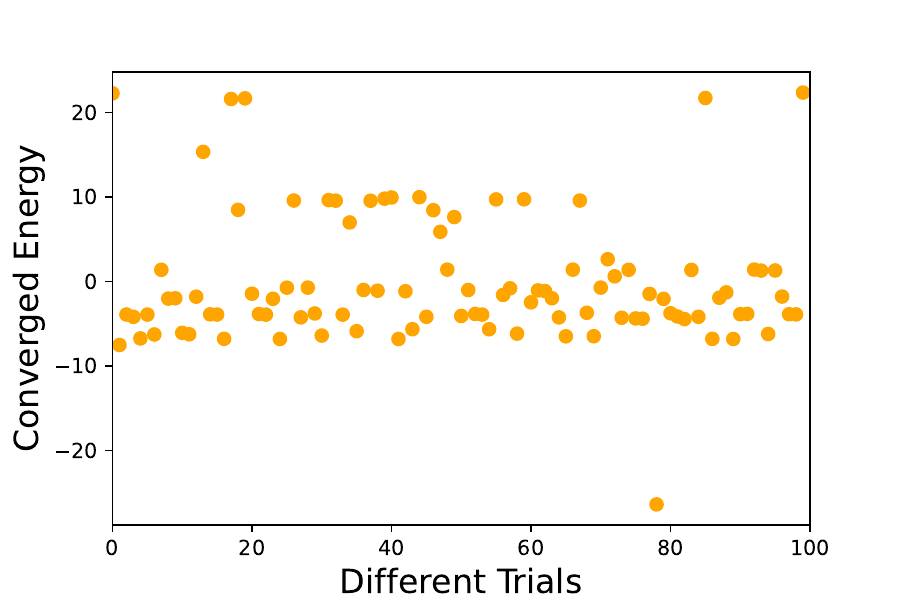}
		\caption{Final converged energy for $W=8.0$ and input state of bad $\theta^{0}$; $N=10$; depth=3.}
		\label{fig:final_10}
	\end{figure}

	Another interesting observation is that EIPR for the input states $\vert \psi_{0} \rangle$, created with the preparation circuit, will decrease as the system size increases as shown in Fig. \ref{fig:eiprW8.0}.  For a larger system, EIPR of thermal phases and MBL phases are both close to $0$. Since our setup in this paper starts from input state with small EIPR on purpose, the approach is still suitable for larger size systems.

	A straightforward question is whether the final converged states of the thermal phase are low-lying excited states that also exhibit ``area law" entanglement, which may compromise the characteristic gap of EIPR between MBL and thermal phases as shown in Fig. \ref{fig:eipr_final}. As mentioned in the adaptive VQE-X algorithm \cite{zhang_adaptive_2021}, we also find a weakly linear relationship between energy of input states and energy of output states. With $\theta^{0}=0.4$, most of the output states are highly excited states that exhibit ``volume law" entanglement (see Fig. \ref{fig:N12W25D12}). So the entanglement difference induces the gap of EIPR between MBL and thermal phases.

	\begin{figure}[t]\centering
	    \setlength{\leftskip}{160pt}
		\includegraphics[width=0.48\textwidth]{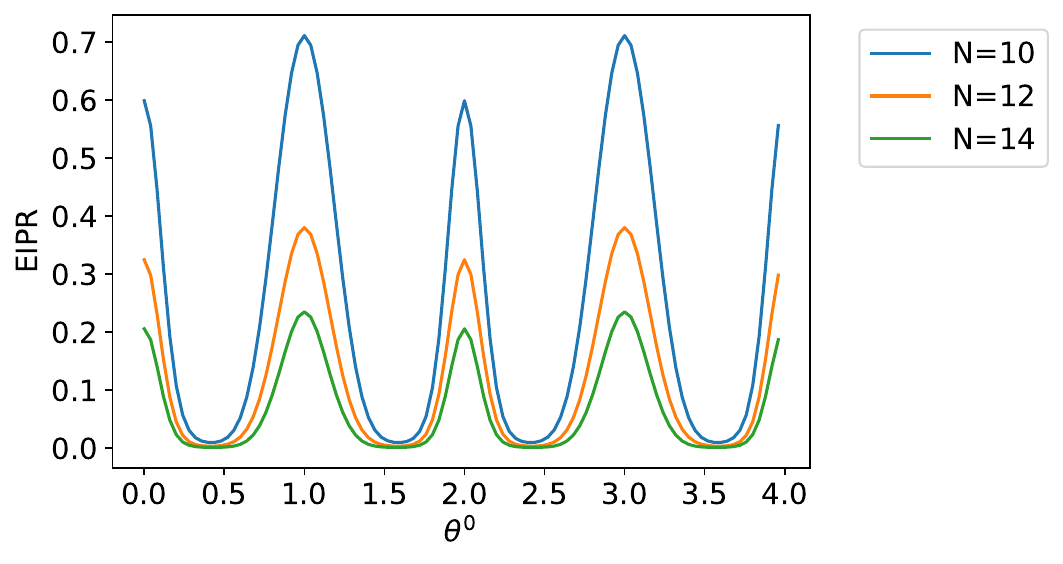}
		\caption{EIPR of input state $\vert \psi_0\rangle$ determined by the preparation circuit with $W=8.0$; $N=10,12,14$.}
		\label{fig:eiprW8.0}
	\end{figure}

	\begin{figure}[t]\centering
		\includegraphics[width=0.40\textwidth]{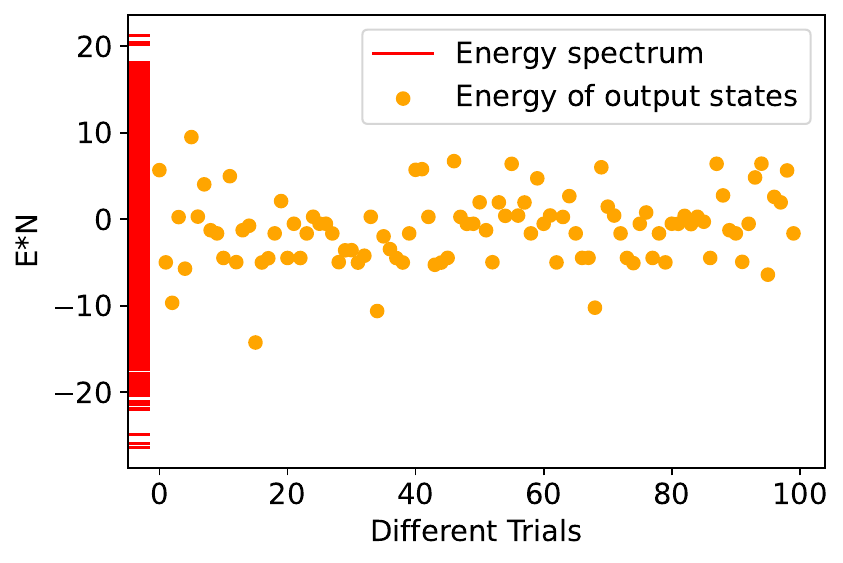}
		\caption{Final output states' energy with $W=2.5$; $N=12$; depth=12. Most of the output states are highly excited states.}
		\label{fig:N12W25D12}
	\end{figure}

	\subsection{D. Scaling analysis of EIPR and eigenstate witness}
	The results of EIPR with $N=12$ in Fig. \ref{fig:eipr_final} show an identifiable gap between MBL and thermal phases. To extend our method to larger size systems, we should study the finite size scaling of this gap. To this end, we define: $\Delta \rm{EIPR} = \rm{EIPR_{MBL}} - \rm{EIPR_{thermal}}$, where $W=4.5$ for the MBL phase and $W=2.5$ for the thermal phase. As the system size increases, we set PQC depth=$N$ that also scales with the system size. As shown in Fig. \ref{fig:scaling}, the gap of EIPR between MBL and thermal phases will be more prominent for larger systems. As shown in the main text, on the MBL side, the optimization depth required for the perfect excited-state convergence is in the order $O(N)$. On the thermal side, the depth required is in the order of $O(2^{N})$. The ideal gap is close to 1 when the optimization for thermal phases completely fails with only depth=$N$ PQC for larger system.

	For current quantum computing hardware, we must consider the effects of the noise. In particular, the noise accumulates with the system size/circuit depth. One important question is whether we can apply our method to a real device with moderate noise. To answer this question, we have also studied the finite size scaling of $\Delta r$: the eigenstate witness difference between the MBL phase ($W=4.5$) and the thermal phase ($W=2.5$). To extend our method to larger size systems, We carry out Monte Carlo trajectory simulation \cite{zhang_tensorcircuit_2022} in the presence of a quantum depolarizing channel with noise strength $p=10^{-3}$ after each two-qubit gate. Quantum channels are added to the circuit by providing a list of their associated Kraus operators, and the mixed state is effectively simulated by an ensemble of pure states. Via this numerical technique, larger size systems are accessible compared to density matrix simulator. As the system size increases, we set PQC depth=$N/2$ that also scales with the system size. As shown in Fig. \ref{fig:scaling_r}, the eigenstate witness $r$ decreases with increasing system size due to stronger effective noise. However, the eigenstate witness difference $\Delta r$ becomes larger when the system size increases, remaining qualitatively the same as the noiseless case shown in Fig. \ref{fig:scaling}. 

	Based on these convincing numerical results, it is reliable to apply our method to larger size systems.

	\begin{figure}[t]\centering
		\includegraphics[width=0.40\textwidth]{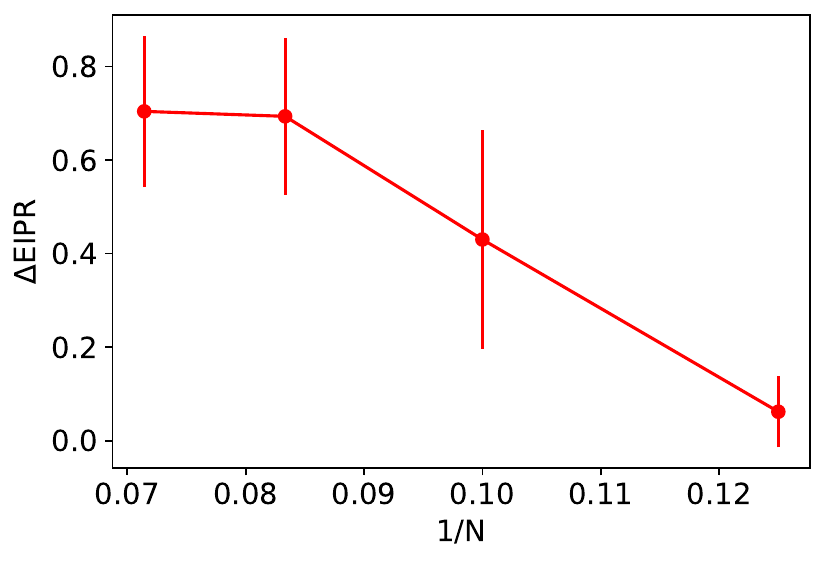}
		\caption{$\Delta \rm{EIPR}$ with $N=8,10,12,14$.}
		\label{fig:scaling}
	\end{figure}

	\begin{figure}[t]\centering
		\includegraphics[width=0.80\textwidth]{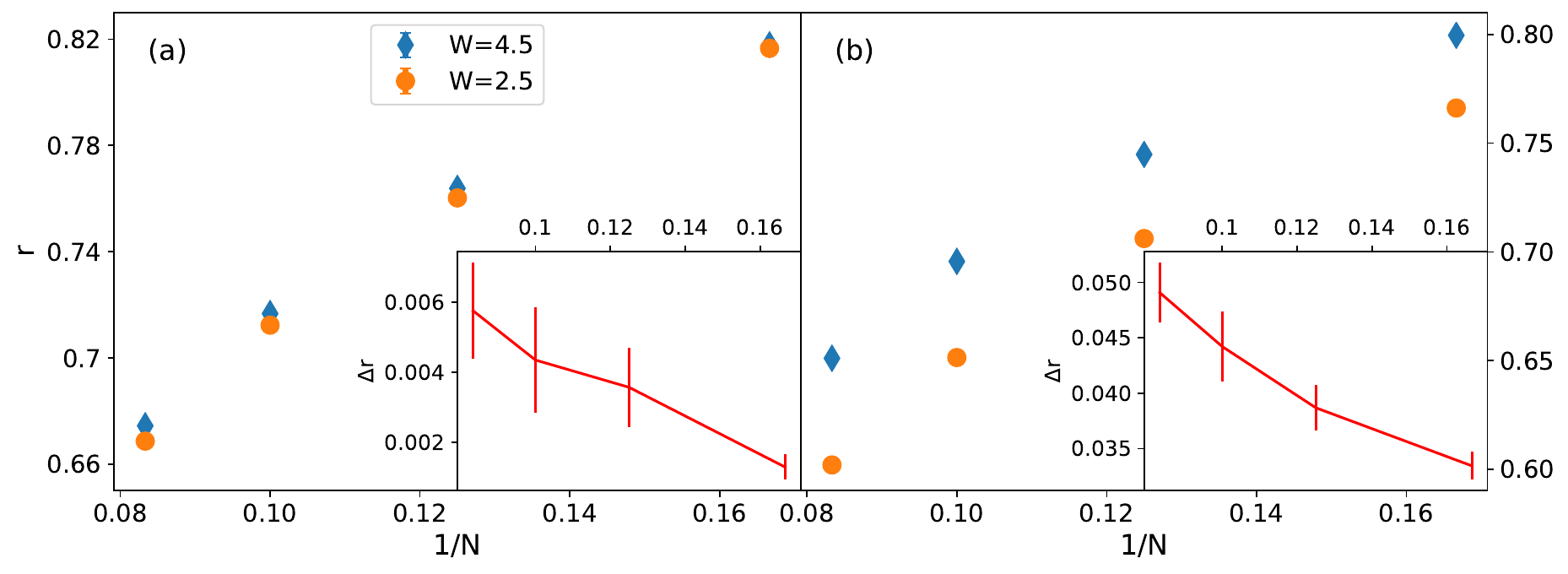}
		\caption{(a), $r$ of final converged states without Trotter decomposition. (b), $r$ of final converged states with one time slice Trotter decomposition. The insets are the eigenstate witness difference $\Delta r$ between the MBL phase and the thermal phase. $N=6,8,10,12$.}
		\label{fig:scaling_r}
	\end{figure}

	\subsection{E. Eigenstate witness}
	The state after the controlled time evolution can be written as:
	\begin{eqnarray}
	\frac{\vert0\rangle \vert\psi\rangle +\vert1\rangle \rm{e}^{-iHt} \vert\psi \rangle}{\sqrt{2}}.
	\end{eqnarray}
	After a partial trace over the system qubit register, the reduced density matrix for the ancilla qubit reads:
	\begin{eqnarray}
	\rho_{\rm{reduced}}=\left(	
	\begin{matrix}
	\frac{1}{2} & \frac{1}{2}\sum_{n}\vert\alpha_{n}\vert^{2}\rm{e}^{i\lambda_{n}t}\\
	\frac{1}{2}\sum_{n}\vert\alpha_{n}\vert^{2}\rm{e}^{-i\lambda_{n}t}& \frac{1}{2}
	\end{matrix}
	\right),
	\end{eqnarray}
	where $\vert\psi\rangle =\sum_{n}\alpha_{n} \vert n\rangle$ such that $H \vert n\rangle = \lambda_{n}\vert n \rangle$. The eigenstate witness is defined as:
	\begin{align}
	r = \rm{Tr}(\rho_{\rm{reduced}}^{2}) &= \frac{1}{2}(\sum_{n}\vert\alpha_{n}\vert^{4}) \\ \no
	&+\frac{1}{2}(1+\sum_{n\neq m } \vert\alpha_{n}\vert^{2} \vert\alpha_{m}\vert^{2} \cos((\lambda_{n}-\lambda_{m})t)) \\ \no
	& \ge \sum_{n}\vert\alpha_{n}\vert^{4} \\ \no
	&= \rm{EIPR}.&
	\end{align}
	We can see the eigenstate witness $r$ is lower bounded by EIPR. For almost all $t$, obtaining an eigenstate witness of $1$ implies that $\vert \psi \rangle$ is an eigenstate of the Hamiltonian $H$. When the evolution time is assumed to be long, $r=\frac{1}{2}(1+\text{EIPR})$; when the evolution time is assumed to be short, i.e. $\text{max}_{m,n}\vert \lambda_{m}-\lambda_{n} \vert t \le \frac{\pi}{2}$, then 
	\begin{eqnarray}
	r \le \frac{1+\cos(\sum_{m,n} \vert \alpha_{m} \vert^2 \vert \alpha_{n} \vert^2  (\lambda_{m}-\lambda_{n})t)}{2},
	\end{eqnarray} 
	given by the Jensen's inequality and the right hand side of which is a function that decreases monotonically with the average eigenvalue gap. Therefore, the eigenstate witness is also a great indicator to assess the converged performance of the excited-state VQE (more details can be found in the Supplemental Material of \cite{santagati_witnessing_2018}).

	In experiments, the eigenstate witness $r$ can be measured using randomized measurements by measuring quantum states in $m$ randomly chosen single qubit basis. We prepare the quantum state $\rho_{\rm{reduced}}$ and rotate it into a random basis with the unitary $V^{(n)}$. Then we measure $s$ samples of the rotated state $V^{(n)}\rho_{\rm{reduced}}V^{(n)^{\dag}}$ in the computational basis and estimate the probability $P^{(n)}(v_{k})$ of measuring the computational basis state $v_{k}$ for the basis $n$. This procedure is repeated for $m$ different measurement bases, and $r$ is then calculated as:
	\begin{eqnarray}
	r=\sum_{v_{k},v_{q}}(-2)^{-D(v_{k},v_{q})} \sum_{n=1}^{m}P^{(n)}(v_{k})P^{(n)}(v_{q}),
	\end{eqnarray}
	where $D(k,q)$ is the Hamming distance that counts the number of bits that differ between the computational states $v_{k}$ and $v_{q}$. Because in our study, we only need to measure the one-qubit density matrix:
	\begin{eqnarray}
	\rho_{\rm{reduced}} =\left(	
	\begin{matrix}
	a     & b\\
    b^{*} & 1-a
	\end{matrix}
	\right),
	\end{eqnarray}
	we can determine the density matrix by the expectation value $\langle x \rangle$, $\langle y \rangle$ and $\langle z \rangle$  after a simple algebraic calculation:
	 \begin{eqnarray}
	 a &=& \frac{\langle z \rangle + 1}{2}, \\
	 b &=& \frac{\langle x \rangle }{2} + \frac{\langle y \rangle }{2}i.
	 \end{eqnarray}

	\subsection{F. Quantum noise and Trotter decomposition}
	In the eigenstate witness measurement case, we have to discuss in terms of density matrix language when noise is considered. After time evolution controlled by ancilla qubit, the reduced density matrix can be written as:
	\begin{eqnarray}
	\rho_{\text{reduced}}=\begin{bmatrix} 1/2 \rm{Tr}(\rho)& 1/2 \rm{Tr}(\rho U^\dagger)\\
	1/2 \rm{Tr}(U\rho)&1/2 \rm{Tr}(U\rho U^\dagger)\end{bmatrix},
	\end{eqnarray}
	where $U=\rm{e}^{iHt}$. Considering the noise added by depolarizing channel, the reduced density matrix is rescaled as:
	\begin{eqnarray}
	\rho_{\text{reduced}} = (1-p_{\rm{eff}})\rho_{\text{reduced}} + p_{\rm{eff}} I/2,
	\end{eqnarray}
	where $p_{\rm{eff}}=33Np$. The coefficient 33 of $p_{\rm{eff}}$ is determined by the number of two-qubit gates in the controlled time evolution circuit. Constructing an optimal quantum circuit for a general two-qubit gate requires at most 3 CNOT gates and 15 elementary one-qubit gates \cite{vatan_optimal_2004}. After adding the control qubit, it requires at most 3 CCNOT gates and 15 two-qubit gates. Each CCNOT gate can further decompose as 6 CNOT gates \cite{mandviwalla_implementing_2018}. So the total number of two-qubit gates is 33. If we also consider the effect of Trotter decomposition, the unitary time evolution matrix can be written as:
	\begin{eqnarray}
	U(t) = \prod_{i}^{N} \rm{e}^{iH_{i,i+1}t},
	\end{eqnarray}
	where $H_{i,i+1}=\sigma^{x}_{i}\sigma^{x}_{i+1}+\sigma^{y}_{i}\sigma^{y}_{i+1}+V_{0}\sigma^{z}_{i}\sigma^{z}_{i+1}+\frac{h_{i}}{2}\sigma^{z}_{i}+\frac{h_{i+1}}{2}\sigma^{z}_{i+1}$, $h_{i} = W \cos(2\pi\eta i +\phi)$. In this part, we can also use a PQC with fewer two-qubit gates to approximate the controlled time evolution and it may further reduce the effects of noise.

	\subsection{G. Details for the real hardware experiments}

	The realization of the controlled time evolution is actually quite challenging on NISQ devices \cite{PhysRevResearch.1.013006, PhysRevA.104.042418}. To demonstrate our method on a four-qubits real hardware, we can use a variational quantum circuit to approximate the controlled time evolution module and reduce the number of two-qubit gates required as discussed in the main text. 

	We need to optimize the parameters of PQC for the excited-state VQE and the controlled time evolution VQC. In our study, we set $\rm{D}_{\rm{VQE}} = 1$ because of the small system size and $\rm{D}_{\rm{VQC}} = 6$ for a better approximation (fidelity around 0.97). Only $\rm{D}_{\rm{VQC}}*N$ ($6*4$) two-qubit gates are required for the controlled time evolution VQC while $33*N$ ($33*4$) two-qubit gates are required for the controlled time evolution with one time slice Trotter decomposition. Actually, there is a tradeoff between the depth $\rm{D}_{\rm{VQC}}$ and the fidelity when using VQC to approximate the controlled time evolution: deeper depth $\rm{D}_{\rm{VQC}}$ can achieve higher fidelity, in the meanwhile, more resources are required. As suggested by the results obtained from noisy simulation and the theoretical understanding of eigenstate witness, we don't need perfect controlled time evolution to discriminate the two phases. The VQC with relatively high fidelity is enough to distinguish the MBL and the thermal phases. Therefore, for larger systems, it is feasible to utilize a VQC to approximate the controlled time evolution with much fewer two-qubit gates than those required in Trotter decomposition.

\end{widetext}

\end{document}